\magnification 1031       
\hsize = 5true in            
\vsize = 6.5true in         
\centerline{\it Preprint from Proceedings}
\centerline{II Int. Symposium on Fundamental Problems in Quantum 
Physics}
\centerline{\it Oviedo, Spain---21-26 July 1996}
\topglue 0.3in              
\leftline{\bf DICHOTOMIC FUNCTIONS AND BELL'S THEOREMS}     
\vglue 0.43in                
\leftline{\bf \hglue 1in A. F. Kracklauer}                  
\vglue 0.125in               
\leftline{\it \hglue 1in Belvederer Allee 23c}              
\leftline{\it \hglue 1in 99425 Weimar, Germany}             
\leftline{\it \hglue 1in kracklau @ fossi.hab-weimar.de}    
\vglue 0.43in                 
\parindent 0in                
It is shown that correlations of dichotomic functions can 
not conform to results from Quantum Mechanics.  Also, it is 
seen that the assumptions attendant to optical tests of Bell's
Inequalities actually are consistent with classical physics so
that in conclusion, Bell's Theorems do not preclude hidden variable 
interpretations of Quantum Mechanics.
\vglue 0.125in               
Key words: Bell's Theorems, Kochen-Specker Theorem, quantum mechanics 
\vglue 0.125in               
\parindent 0.4in             
The analysis attending Bell's Theorems imposes on hidden variable 
formulations of Quantum Mechanics (QM) the requirement that they be 
able to correlate dichotomic functions (representing spin) so as to  
duplicate the result from QM:
$$P_{\bf a,b}=\int\rho (\lambda )\,A({\bf  a},\lambda )B({\bf  b,\lambda 
)}d\lambda,=-\cos({\bf \,a,b}),\,\eqno(1)$$
where content and notation are taken from Bell.[1]

Note, however, that {\it inter alia}, only the right side of Eq. (1) 
is a harmonic function.  This means that the equation itself is an 
absurdity.  To see this, consider the derivative of the QM expectation 
expressed as a correlation of dichotomic functions, $D_{a,b}$ which 
change sign at the points $x_{j}$ ($\delta$ is the Dirac delta 
function): $$\eqalignno{ -{\partial cos(\tau)\over \partial\tau} & 
\propto {\partial \over\partial \tau } {1\over T} \int^{T}_{0}D_{a}(x-
\tau) D_{b}(x)dx, \cr sin(\tau) & \propto  {1\over T}\int_{0}^{T} 
\sum_{n}^{N}\pm\delta(x_{n})D_{b}(x)dx, \cr {\phantom x} \, &\propto  
{1\over T}\sum_{n}^{N}(\pm)D_{b}(x_{n})\, = {integer \over T}. &(2) 
\cr}$$ (The sign of individual terms in the above series depends on 
the specific form of $D_{a,b}$, which, for present purposes, is not 
required.) A further differentiation by $\tau$ yields $cos(\tau)\equiv 
0,$ a false equation.  

Nothing was in play here but the dichotomic nature of  the functions 
$D$; it is clear that such functions simply can not be correlated to 
yield harmonic functions.[2]  Obviously, if dichotomic functions 
cannot yield the QM result, then QM is doing something other than 
correlating them.  If this be the case, then a hidden variable 
alternative to QM need not be required to so correlate them and the 
logic of Bell's Theorem is broken. 

Likewise, the famous result from Greenburger et al.[3] showing that 
the correlation of three or more dichotomic functions representing 
three or more particles is never; i. e., even for perfect correlations, 
compatible with the harmonic functions given by QM, is attributable to 
the same structural cause. This is most easily seen by expressing 
multiple correlations (for three particles, say) in terms of 
dichotomic sequences $A, B, C...$ (i. e., data points) as follows 
$$-\cos(\tau + \theta) = {1 \over{NM}} \sum_{j,k}^{N,M} 
A_{j,k}B_{j,k}(\tau)C_{j,k}(\theta), \eqno(3)$$ 
where (ignoring for the moment obvious analytic absurdities) it is 
clear that even for $\tau + \theta = 0$ or $\pi$; i. e., for perfect 
(anti) correlations, the equation can not hold because odd (even) 
multiples of $-1$ do not (do) cancel.  This exposes an internal 
contradiction in no need of empirical verification. (Degenerate cases 
occurring for even multiples of dichotomic functions do not invalidate 
the principle. Moreover, for those with strong intuition, it is clear 
that  ``n-chotomic'' functions of any rank will result in analytic 
problems.)

The question arises: if QM is not doing the mathematically impossible, 
what is it doing?  For the case of polarized ``photons,'' used 
virtually exclusively for testing Bell's Inequalities, QM is in fact 
giving a classical result. The analysis of optical analogues of the 
EPR event artificially associates an unpolarized with an 
anticorrelated state by {\it defining} [4] the  ``coefficient of 
correlation'' to be 
$$E({\bf a,b})\equiv P_{(++)}+P_{(--)}-P_{(+-)}-
P_{(-+)}, \eqno(4)$$ 
where the $P's $ are the ``quantum mechanical predictions'' for 
detection coincidences 
$$\eqalignno{ P_{(++)}=P_{(--)}=&{1\over 2}\cos^2({\bf a,b}),\cr  
P_{(+-)}=P_{(-+)}=&{1\over 2}\sin^2({\bf a,b}), &(5)\cr}$$ 
so that Eq.~(4) becomes 
$$E_{\bf Optical}=\cos2({\bf a,b}). \eqno{(6)}$$
Note, however, that the $P'$s just give the intensity of polarized 
light as measured with respect to an axis different from the axis of 
polarization according to Malus' Law---a non quantum rule.  Of course, 
the $P$'s must be suitably interpreted to correspond ultimately to 
``click'' probabilities in experiments performed at minimal intensity 
where the statistics within each polarization state may require more 
than traditional physics. 

This is in total accord with the fact that creation/annihilation 
operators for photons of different states of polarization commute.  As 
is well known, QM differs from classical physics where and only where 
conjugate operators do not commute.  (This observation conforms
with work reported elsewhere in these proceedings[5] wherein supposed 
QM polarization (or spin) correlations were obtained from a model 
involving rubber bands or ``random gun'' and no Planck's constant!) 
Thus, tests to plumb the ontological implications of QM; e. g., 
nonlocality, can not be made on the basis of optical experiments 
involving polarization states.  As the creation/annihilation operators 
for photons of different colors also commute, these arguments extend 
directly to experiments exploiting parametric down conversion and so 
on. 

QM multiparticle correlations, such as Eq. (3) or its  n-particle 
generalizations, as noted elsewhere[6], make contact with the 
Bell-Ko\-chen-Specker Theorem.  Although analysis of such equations, 
restricted as it is to considering perfect (anti-) correlations, is 
less general than the full Theorem, it is more transparent because it 
is just tracking factors of $-1.$  In addition, it is here very clear 
that dependence on variables in addition to $\tau$ and $\theta$; e. 
g., by investing $A, B, C$ with nonlocality, can not circumvent  
inconsistency which results only from their dichotomic values.  Again, 
however, Eq. (3) can be rendered rigorously correct and physically 
sensible in any n-particle generalization if dichotomic sequences $A, 
B, C$ etc. are replaced by cosines and sums converted to integrals; i. 
e., by considering classical polarization correlations.      

In conclusion, it is seen that a demand that hidden variable theories 
should correlate dichotomic functions to get a harmonic result is ill 
founded.  When this stipulation is relaxed, then the extraction of 
Bell's inequalities does not go through and therefore hidden variable 
theories duplicating QM polarization correlations; e. g., [7], are not 
precluded in principle. Moreover, because the mathematics describing 
spin correlations; i. e., their interstate structure, is isomorphic to 
that for classical polarization, the underlying phenomena also must be 
classical in essence. 

In other words, it is clear that Bell's Theorems establish beyond 
dispute that self consistent, intuitively clear hidden variable 
theories can not duplicate all of QM.  Indeed such alternatives will 
be (and should be) unable to replicate those very aspects of QM that 
have been the source of confusion and contest from the beginning, 
namely the measurement projection hypothesis and an under restrictive 
identification of vectors in relevant Hilbert spaces as physically 
realizable states; e. g., pure and ``cat'' states and even 
``photons.''  These two features of orthodox QM lead not only to the 
logical contradictions exposed by Bell's Theorems but appear 
essentially untestable in all but certain ``n-chotomic'' circumstances 
where a classical explanation fits anyway.  These observations are 
not here unique.  Many roads lead away from  ``Copenhagen,'' another 
such, for example, ends in LA.[8] 

\vglue 0.125in             
\leftline{\bf REFERENCES}                        
\vglue 0.125IN             
\parindent 0.2in           
\item{1.}J. S. Bell, {\it Speakable  and unspeakable in quantum  
  mechanics}, (Cambridge University Press, Cambridge, 1987).  

\item{2.} P. Claverie and S. Diner, {\it Israeli J. Chem.} {\bf 19}, 
54 (1980), make the same observation in a different context. (I thank 
professors A. M. Cetto and L. de la Pe\~na for this reference.) 

\item{3.}D. M. Greenburger, M. A. Horne, A. Shimony and A. 
Zeilinger, {\it Am. J. Phys.} {\bf 58}(12), 1131 (1990).

\item{4.}A. Aspect, P. Grangier, {\it Proc. Int. Symp. Found. of QM}, 
(Tokyo, 1983), pp.214-224.

\item{5.}B. Coecke, B. D'Hooghe and F. Valckenborgh, {\it These 
Proceedings,} xx(1996). 

\item{6.}M. D. Mermin, {\it Rev. Mod. Phys.} {\bf 65}(2), 803 (1993).

\item{7.}A. O. Barut in {\it Waves and Particles in Light and 
Matter}, Edited by A. van der Merve aUd A. Garuccio (Plenum Press, New 
York 1994), pp. 9-18. 

\item{8.}E. J. Post, {\it Quantum Reprograming}, (Kluwer Academic 
Publishers, Dordrecht, 1995).

\end